# Strain engineering Dirac surface states in heteroepitaxial topological crystalline insulator thin films

Ilija Zeljkovic[1], Daniel Walkup[1], Badih Assaf[1], Kane L Scipioni[1], R. Sankar[2], Fangcheng Chou[2] and Vidya Madhavan[1,3]

[1]Department of Physics, Boston College, Chestnut Hill, Massachusetts 02467, USA [2]Center for Condensed Matter Sciences, National Taiwan University, Taipei 10617, Taiwan [3]Department of Physics and Frederick Seitz Materials Research Laboratory, University of Illinois Urbana-Champaign, Urbana, Illinois 61801, USA

**In newly discovered topological crystalline insulators (TCIs) (Ref. 1), the unique crystalline protection of the surface state (SS) band structure has led to a series of intriguing predictions of strain generated phenomena, from the appearance of pseudo-magnetic fields and helical flat bands [2], to the tunability of the Dirac SS by strain that may be used to construct `straintronic' nanoswitches [3]. However, practical realization of this exotic phenomenology via strain engineering is experimentally challenging and is yet to be achieved. In this work, we have designed an experiment to not only generate and measure strain locally, but to also directly measure the resulting effects on the Dirac SS. We grow heteroepitaxial thin films of TCI SnTe *in-situ* and measure them by using high-resolution scanning tunneling microscopy (STM). Large STM images were analyzed to determine picoscale changes in the atomic positions which reveal regions of both tensile and compressive strain. Simultaneous Fourier-transform STM was then used to determine the effects of strain on the Dirac electrons. We find that strain continuously tunes the momentum space position of the Dirac points, consistent with theoretical predictions [2,3]. Our work demonstrates the fundamental mechanism necessary for using TCIs in strain-based applications, and establishes these systems as highly tunable platforms for nanodevices.**

Controllable manipulation of materials at the atomic length scales is necessary for realizing the next generation of nanodevices. A promising pathway involves the application of strain, which can be used to tune the inter-atomic lattice spacing and induce an accompanying change in the electronic structure. In superconductors for example, strain has been utilized to enhance the superconducting critical temperature [4,5] or to induce resistivity anisotropy [6]. In topological Dirac materials, theoretical studies indicate that both tensile and compressive strain can profoundly alter the characteristics of the topological SS [2,3,7]. However, the experimental realization of large enough strain and the detection of any resulting SS changes have proven difficult. Although recent experiments reported that strain in the vicinity of atomic dislocations in $Bi_2Se_3$ thin films changed the local density of states (LDOS) (Ref. 8), by simply measuring the LDOS, it is nearly impossible to experimentally disentangle the effect of strain from other complicating factors such as charged grain boundaries [9] and topological modes [10,11]. Furthermore, the spatial distribution of these dislocations is difficult to control, which may be one of the main obstacles in their applicability in nanotechnology. For these reasons, exact experimental quantification of the effects of strain on the topological band structure and realization of suitable platforms for applications in the budding area of `straintronics' remain elusive.

TCIs are a recently discovered subclass of materials within the realm of topological Dirac systems. Despite being trivial under the $Z_2$ topological classification [12–14] as a result of even number of band crossings at the Fermi level, these systems harbor topological SS spanning the bulk bands due to the underlying crystal symmetry [1]. In contrast to time-reversal symmetry protected conventional $Z_2$ topological insulators where the Dirac node is pinned to time-reversal invariant (TRI) points [12–14], crystalline symmetry protection in TCIs allows a unique tunability of Dirac points, whose momentum position with respect to the TRI points can be modulated. This momentum shift can be achieved by either changing the alloying composition of the system [15], or by the application of strain [2,3]. Strain therefore presents a particularly intriguing tuning `knob' for engineering the Dirac SS band structure, as it can be used as an *in-situ* equivalent of chemical composition change. Interestingly, the effect of strain on the SS of TCIs goes a step further. Similar to graphene [16–18], strain in TCIs has been predicted to generate large pseudomagnetic fields [2]. These predicted phenomena make the realization of strained TCIs an interesting frontier in the study of Dirac systems.

To achieve this, we designed an experiment to not only induce and measure strain locally, but to also directly correlate it with its effects on the Dirac dispersion. First, to generate strain we utilize the lattice misfit between a TCI SnTe and a trivial bulk insulator PbSe single crystal which is used as the substrate for the growth of SnTe thin films. This is an ideal system since the lattice constant mismatch of ~3% at room-temperature [19,20] between the two is large enough to induce a periodic, two-dimensional structural buckling [19], but also sufficiently small to allow a uniform growth of heteroepitaxial thin films along the desired (001) crystalline direction. Next, by utilizing high-resolution imaging and spectroscopy we were able to measure the small local changes in the lattice constant that generate the strain and show how both tensile and compressive strain can tune the band structure and shift the momentum space location of Dirac nodes.

Thin films were made *in-situ* in an ultra-high vacuum system attached to the STM. SnTe thin films were deposited at 300 °C using the method of electron-beam evaporation at the growth rate of ~ 2 monolayers (ML) per minute for a total ~40 ML thickness as determined by the thickness of SnTe thin films grown on Si(001) under equivalent conditions. The substrates were PbSe single crystals grown by the self-selecting vapor growth method and cleaved at room-temperature in ultra-high vacuum along the (001) direction to expose a pristine surface. Consistent with previous experiments on IV-VI semiconductor heteroepitaxial films [19], STM topographs of our (001) SnTe thin films display the characteristic `checkerboard' buckling nanopattern with the period of 14 +/- 1 nm (Fig. 1(a)). This pattern is present across the field-of-view in all regions of the sample imaged separated by tens of micrometers (*Supplementary Information I*), which demonstrates the uniformity of our growth process. Average *dI/dV* spectra obtained on the surface of these strained SnTe thin film suggest *p*-doping of the system and reveal the minimum in the density of states (approximate energy of the Dirac point) to be around ~300 meV above the Fermi level (Fig. 1(b)). These properties are qualitatively similar to those obtained on SnTe thin films in which no substrate-induced strain is present [21], likely due to a similar defect density.

While the approximate energy of the Dirac point with respect to the Fermi level can be obtained from average *dI/dV* STM spectra obtained over large defect-free areas, details of the SS band structure energy dispersion are more conventionally measured using the method of FT quasiparticle interference (QPI) (Ref. 22). This widely applicable imaging method detects the standing wave `ripples' in two-dimensional *dI/dV* maps arising from the elastic scattering of quasiparticles on the surface of a material, and uses their energy dispersion to extract the band structure of the system. Based on previous angle-resolved photoemission spectroscopy (ARPES) experiments on SnTe single crystals [23,24] and theoretical calculations [25,26], we can schematically represent the constant energy contours (CECs) of the (001) surface far below the Dirac point as four hole pockets centered at each X point within the first Brillouin zone (BZ), symmetric around (110) and (-110) crystalline directions (Fig. 2(a)). As the first approximation to what scattering signature we might expect, disregarding any spin or orbital matrix elements, it is sufficient to take an autocorrelation of the CECs (Fig. 2(b)). In this system, we will focus on two main sets of inequivalent scattering wavevectors: $Q_1$, which corresponds to the scattering between the mirror-symmetric hole pockets around X and X', and $Q_2$, which represents the scattering between the pockets around X and Y points.

To measure the QPI patterns in our SnTe thin films, we calculate the FTs of experimental *dI/dV* conductance maps acquired over a clean 1300 Å square region of the sample shown in Fig. 3(a) (see *Methods* for more details). Our data (Fig. 2(c)) beautifully reproduces the main scattering modes seen in Fig. 2(b). Zoom-ins of $Q_1$ and $Q_2$ clearly show a change in the peak position with energy (Fig. 2(d)), which is directly attributed to the shrinkage of the CECs with increased energy within the regime below the Dirac point. Furthermore, based on our QPI measurements of $Q_1$, we can directly visualize higher anisotropy of the CECs in SnTe compared to its TCI cousin $Pb_{0.63}Sn_{0.37}Se$ (see *Supplementary Information II*). This asymmetry is reflected in the rounded-square shape of $Q_2$ in this system (Figs. 2(c,d)), which contrasts the nearly circular feature at the same scattering wave vector in more isotropic $Pb_{0.63}Sn_{0.37}Se$ [27], and is likely rooted in the larger momentum space distance between neighboring Dirac nodes in SnTe compared to that in $Pb_{0.63}Sn_{0.37}Se$ (Refs. 26,27). To quantify the dispersion of scattering wavevectors, we obtain the position of each QPI peak by Gaussian fitting, and plot its dispersion as a function of energy (Figs. 2(e,f)). The dispersion of $Q_1$ is directly proportional the SS dispersion along the Γ-X direction, and as expected for topological Dirac systems, shows monotonic, nearly linear energy dependence. Moreover, the dispersion of the more prominent peak $Q_2$ allows us to determine an approximate position of the Dirac point of ~0.06 Å$^{-1}$ away from X, consistent with the theoretically expected value of 0.05 Å$^{-1}$ (Ref. 26) (*Supplementary Information II*).

Having understood the electron scattering signature and its relation to the SS band structure, we proceed to determine the lattice strain induced by the `checkerboard' structural buckling. STM topographs demonstrate both the periodic nature of this buckling as well as the high-quality of our samples (Fig. 3(a)). The `checkerboard' corrugation amplitude of ~1 Å (Fig. 3(c)) remains unchanged as a function of bias polarity used (*Supplementary Information III*), signaling that the buckling observed is purely of structural origin. In order to quantify the local lattice distortion, we apply the Lawler-Fujita algorithm [28] on an atomically-resolved STM topograph. The purpose of this algorithm is to detect the shift of atoms from their ideal square-lattice positions with picometer precision. Using this information, we can construct a two-dimensional `strain map' corresponding to the topograph (Fig. 3(b)), which allows us to visualize the spatial evolution of strain at the atomic length scales (*Supplementary Information III*). The measured strain evolves smoothly across the sample (Figs. 3(d)), and exhibits an excellent correlation with the topographic buckling (correlation coefficient between the two is ~0.85). Positive values (red regions) in Fig. 3(b) denote tensile strain in the elevated topographic areas, while the negative values (blue regions) denote compressive strain in the topographic valleys (troughs). We note here that while the STM topograph height variations could be used to estimate strain, the height measured by STM is a complex function of the actual topograph, density of states variations and tip condition and does not therefore provide an accurate quantitative measure of strain. Our method to measure strain however provides not only qualitative information on the strain variation, but also quantitative information on the actual strain magnitude.

As the next step towards determining how spatially varying strain affects the electronic band structure, we utilize the following approach. We first separate each *dI/dV* real-space conductance map used for band structure mapping in Figure 2 into three different regions of equal areas based on the topographic corrugations: (1) region of topographic elevations (tensile strain), (2) region of topographic valleys (compressive strain), and (3) the remaining region (corresponding to approximately zero relative strain) which will be used as a control group (Fig. 4(a), *Supplementary Information III*). Next, we apply FT to each one of the three strain-separated *dI/dV* regions, and compare the resulting FT images (Fig. 4(b-e)). Visual comparison between FTs of regions of tensile (left) and compressive (right) strain suggests a prominent change in the *q*-space positon of the QPI peaks (Fig. 4(b-c)), which is further illustrated by the pronounced difference in the peak positions in the FT line cuts (Figs. 4(d-e)).

To quantify these effects, we plot the energy dispersion of the $Q_1$ peak for the regions with tensile and compressive strain, respectively (Fig. 4(f)). The dispersions reveal a systematic shift of the peak positions between the two regions. We immediately rule out the possibility that this relative shift is the effect of the chemical potential variation since the average Dirac point in the two regions is at the same energy. This can be clearly seen by comparing the average *dI/dV* spectra acquired over different strained regions, which maintain the same shape as well as position of the minimum in the density of states, which based on the band structure revealed by ARPES [23, 24] is expected to represent the energy of the Dirac point (Fig. 1(b)) or at the least lie very close to it. To explain the relative shift, we turn to theoretical calculations of the effects of strain. Theoretically, compressive strain is predicted to cause a shift of the Dirac points in momentum space towards Γ, whereas the tensile strain should induce the shift of Dirac points towards X (Refs. 2,3) as schematically shown in Figure 4(h). Our data beautifully demonstrate this picture – compressive (tensile) strain, generates a momentum shift in the dispersion along the Γ-X direction towards the Γ (X) point (Fig. 4(f)). Furthermore, based on our measurements, we are able to calculate the magnitude of the shift to be ~0.015 Å$^{-1}$ per one percent of the lattice distortion, comparable to the theoretically obtained value of ~0.023 Å$^{-1}$ for the equivalent structural distortion magnitude in a related material PbTe [2].

Our work paves the way towards strain engineering of the topological SS band structure in these and other related materials. For example, in thin films of IV-VI semiconductors PbSe and PbTe, relatively small strain of less than 4% is expected to induce a crossover between the topological and the trivial regime [3], which could be used in creating strain-controlled nanoswitches in quantum devices. We also note that although the local strain in our thin films has not been sufficient to generate the pseudomagnetic Landau levels (LLs) predicted by theory [2], the pseudomagnetic LLs might still appear in thin films of reduced thickness or in those grown on a substrate with larger lattice mismatch, since both of these pathways would increases the lattice strain at the surface. Our experiments provide a platform for realizing these exotic phenomena.

## Methods

All d$I$/d$V$ measurements were acquired at ~4.5 K using a standard lock-in technique with ~10 meV peak-to-peak modulation and 1488 Hz frequency. We use Lawler-Fujita drift-correction algorithm [28] on all acquired data to remove the effects of slow thermal and piezoelectric drift.

# Figures

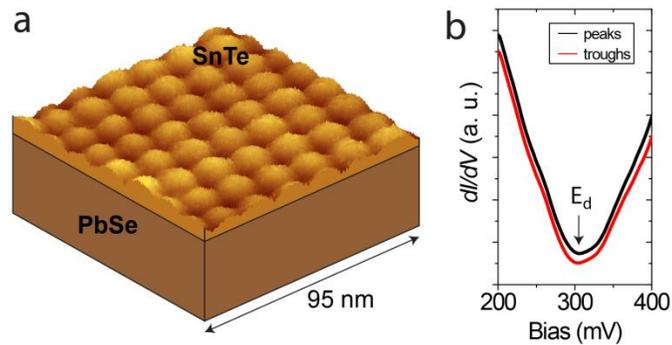

**Figure 1.** SnTe heteroepitaxial thin films. (a) Schematic of the SnTe thin film grown on cleaved PbSe single crystal substrate. Due to lattice mismatch between SnTe and PbSe, a two-dimensional structural buckling pattern is formed at the surface of the thin film. (b) Average *dI/dV* spectra acquired in the elevated topographic regions (peaks; black) and the valleys (troughs; red). Other than a relative intensity shift due to the setup condition, no relative shift in energy between the spectra is observed, which rules out the effects of chemical potential variation between the two distinct topographic regions. Setup conditions in: (a) $I_{set}$=100 pA, $V_{set}$=-50 mV; (b) $I_{set}$=350 pA, $V_{set}$=200 mV.

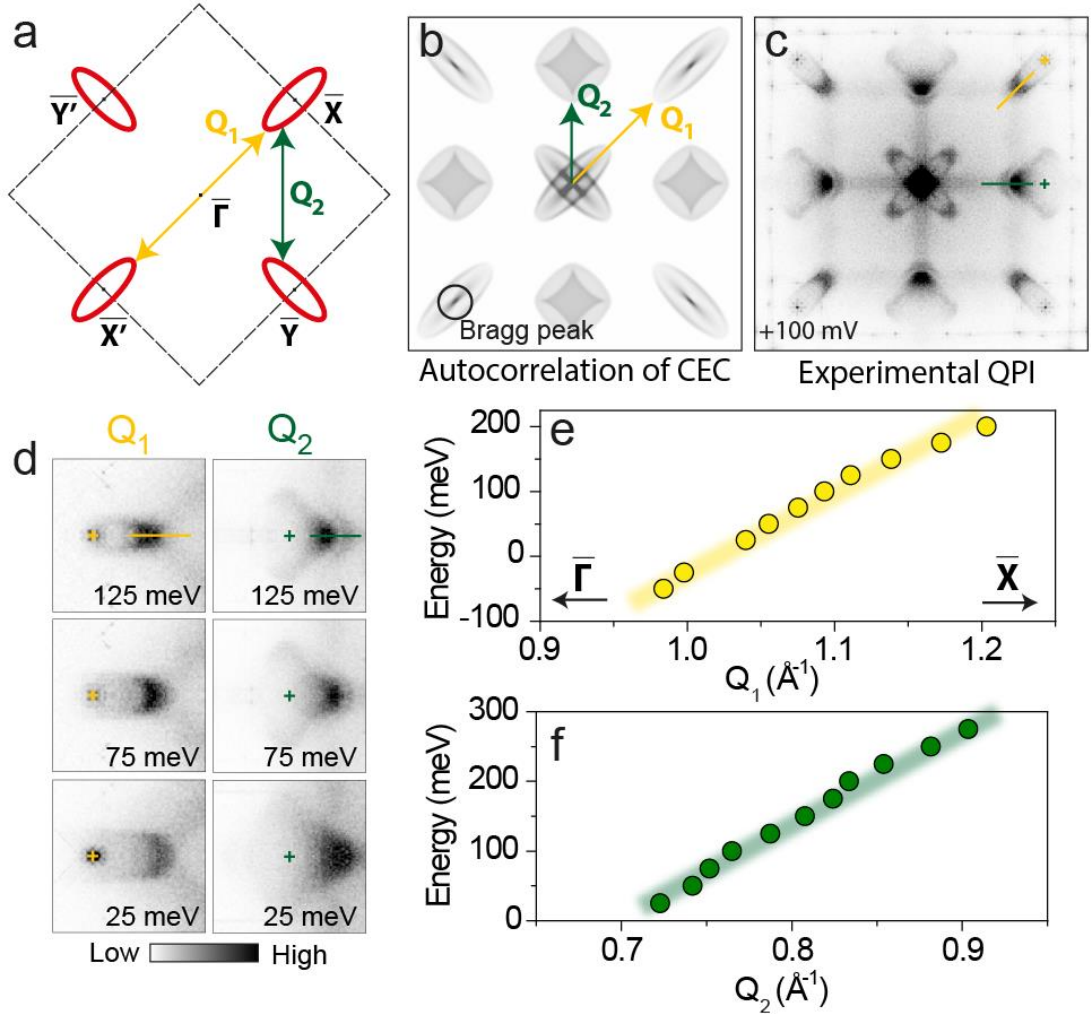

**Figure 2.** Quasiparticle interference (QPI) measurements. (a) Schematic of the constant-energy contours (CECs) below the lower Lifshitz transition, consisting of four hole pockets (red curves) around each X point. Dashed lines denote the first BZ. (b) Autocorrelation of the CECs in (a) showing the two main scattering wavevectors $Q_1$ and $Q_2$. (c) FT of experimental $dI/dV$ conductance image acquired at +100 mV showing an excellent match with the expected QPI pattern in (b). (d) Zoom-in on the two main sets of scattering wave vectors: $Q_1$ (left column) and $Q_2$ (right column) for several representative energies. Green and yellow crosses in (d) represent the same positions in $q$-space as those denoted in (c); green and yellow lines in (d) represent the orientation of the FTs also shown in (c). Energy dispersion of (e) $Q_1$ and (f) $Q_2$. Thick, diffuse green and yellow lines in (e,f) are visual guides. All experimental $dI/dV$ maps have been 1-pixel boxcar averaged in $q$-space (effective smoothing radius of ~0.5% of the atomic Bragg peak wave vector) and four-fold symmetrized.

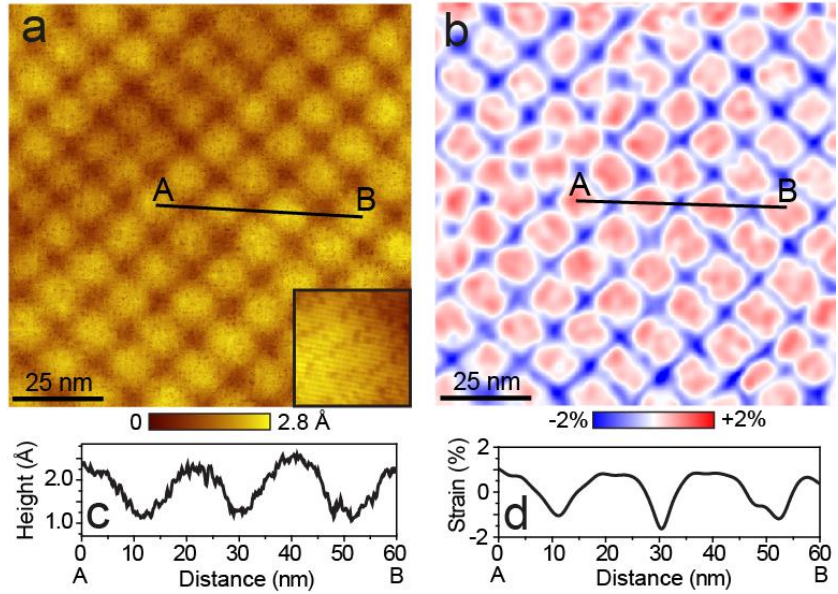

**Figure 3.** Mapping the local lattice strain. (a) STM topograph of the region of the sample used for QPI measurements in Figure 2. Atomically-resolved inset in (a) acquired over a 10 nm square region demonstrates the high quality of our samples. (b) Strain map calculated from the measured local atomic positions in the topograph in (a). Positive values (red) represent tensile strain (increased lattice constant), while the negative values (blue) denote the compressive strain (decreases lattice constant). (c) Line cut through topograph in (a) between points A and B defined in (a) which allows us to visualize ~1 Å topographic corrugations due to structural buckling. (d) Line cut through strain map in (b) between two points defined in (b). STM setup conditions in: (a) $I_{set}$=200 pA, $V_{set}$=-100 mV; inset in (a): $I_{set}$=100 pA, $V_{set}$=-200 mV.

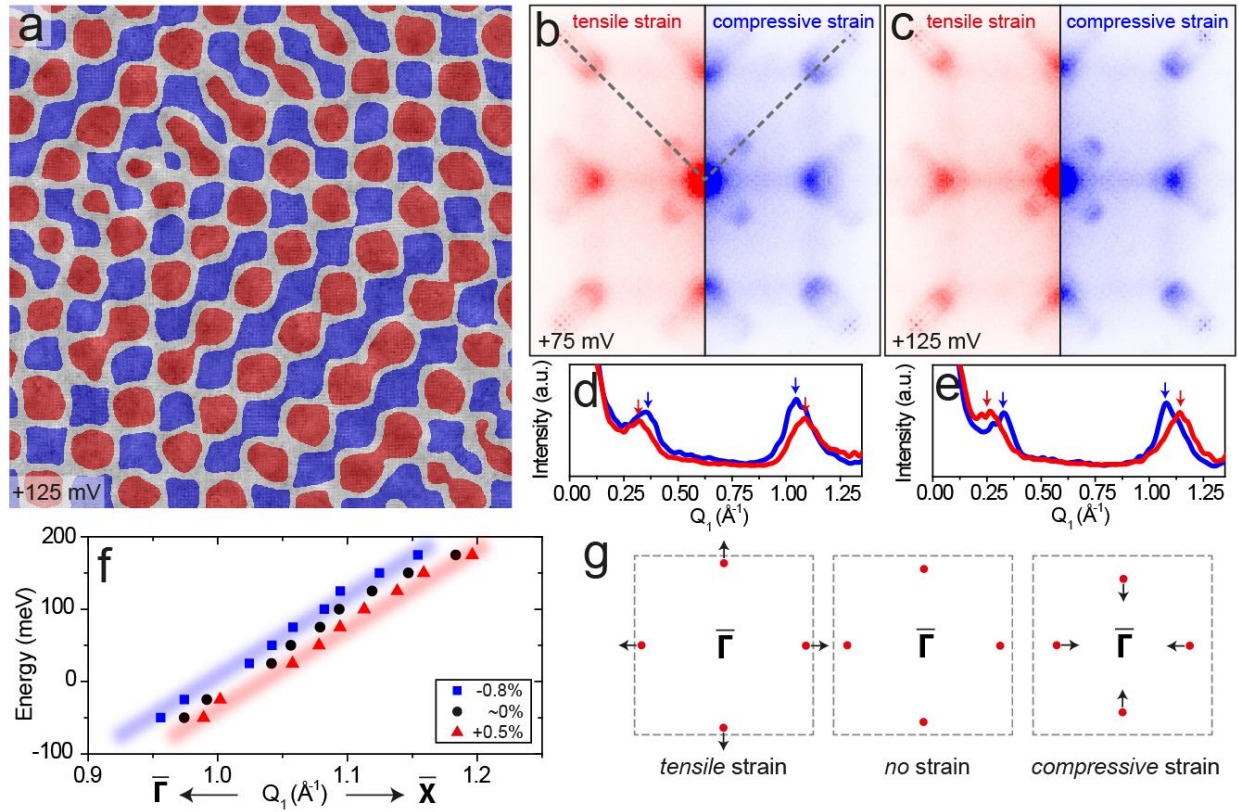

**Figure 4.** Strain engineering of the SS band structure. (a) *dI/dV* map at 125 mV color coded to denote tensile strain (red), compressive strain (blue) and no relative strain (light gray). (b-c) Experimental FTs of *dI/dV* conductance maps masked by the magnitude of strain: tensile strain of 0.5% (left; red) and compressive strain of -0.8% (right; blue). The FTs have been four fold symmetrized. Pronounced difference in the peak positions can be observed by eye. (d-e) Line cuts through the FTs in (b-c) along the lines denoted in (b) from the center of the FT to the Bragg peak. Both peaks within a single blue curve are closer together compared to the same peaks in the corresponding red curve, emphasizing the change in the SS band structure. (f) $Q_1$ dispersion along the Γ-X direction for +0.5% tensile strain (red triangles), approximately zero relative strain (less than 0.1%; black circles) and -0.8% compressive strain (blue squares). The three curves are clearly offset in momentum. (g) Schematic representation of the expected band structure change with the applied strain. Red circles represent the positions of the Dirac nodes, dashed square denotes the first BZ, and arrows represent the direction of the Dirac node shift. Under tensile strain (left) Dirac points move away from the Γ point compared to the same Dirac points when no strain is applied (middle). Equivalently, under compressive strain (right), the Dirac nodes move towards the Γ point. Experimental SS dispersions in (f) clearly demonstrate the schematic in (g).